\documentclass[10pt,twocolumn,twoside]{IEEEtran}

\IEEEoverridecommandlockouts

\usepackage{graphicx,psfrag}
\usepackage{amsmath,amssymb,amsfonts,mathrsfs}
\usepackage{color,epstopdf}
\usepackage{algorithmic}

\newtheorem{remark}{Remark}
\newtheorem{assumption}{Assumption}

\newtheorem{theorem}{Theorem}
\newtheorem{lemma}{Lemma}

\newtheorem{proposition}{Proposition}

\newcommand\myrule{\hrule width \columnwidth height .4pt}

\newcommand{\until}[1]{\{1,\dots, #1\}}
\newcommand{\subscr}[2]{#1_{\textup{#2}}}

\newcommand{\setdef}[2]{\{#1 \; : \; #2\}}

\newcommand{\map}[3]{#1: #2 \rightarrow #3}

\newcommand{\integernonnegative}{\ensuremath{\mathbb{Z}}_{\ge 0}}
\newcommand{\real}{\ensuremath{\mathbb{R}}}
\newcommand{\realpositive}{\ensuremath{\real_{>0}}}
\newcommand{\realnonnegative}{\ensuremath{\real}_{\geq0}}

\newcommand{\R}{\mathbb{R}}

\newcommand{\eps}{\varepsilon}
\newcommand{\E}{\mathcal{E}}    

\newcommand{\neigh}[1]{{\cal N}_{#1}}

\newcommand{\degmax}{\subscr{d}{max}}
\renewcommand{\deg}{d}

\newcommand{\sign}{\operatorname{sign}} %

\newcommand{\dst}{\displaystyle}
\newcommand{\degmin}{\subscr{d}{min}}

\def\qedp{\hspace*{\fill}~{\tiny $\blacksquare$}}

\def\be{\begin{equation}}
\def\ee{\end{equation}}
\def\ba{\begin{array}}
\def\ea{\end{array}}
\def\eqa{\begin{eqnarray}}
\def\eqe{\end{eqnarray}}

\def\stopmodif{\color{black}}

\begin{document}
\title{A Jamming-resilient Algorithm for Self-triggered Network Coordination}

\author{Danial~Senejohnny,~
        Pietro~Tesi,~
        and~Claudio~De~Persis 
\thanks{Danial Senejohnny, P. Tesi, and C. De Persis are with ENTEG and Jan C. Willems Center for Systems and Control, University of Groningen, 9747 AG Groningen, The Netherlands e-mail: \{d.senejohnny, p.tesi, c.de.persis\}@rug.nl.}
}


\markboth{
}{Senejohnny \MakeLowercase{\textit{et al.}}: A Jamming-resilient Algorithm for Self-triggered Network Coordination}


\maketitle

\begin{abstract}
The issue of cyber-security has become ever more prevalent in the analysis and design of cyber-physical systems. 
In this paper, we investigate self-triggered consensus networks in the presence of communication failures 
caused by Denial-of-Service (DoS) attacks. A general framework is considered in which 
the network links can fail independent of each other.
By introducing a notion of Persistency-of-Communication (PoC), we provide an explicit 
characterization of DoS frequency and duration under which consensus can be preserved
by suitably designing time-varying control and communication policies.
An explicit characterization of the effects of DoS on the consensus time is also provided.
The considered notion of PoC is compared with classic average connectivity 
conditions that are found in pure continuous-time consensus networks. 
Finally, examples are given to substantiate the analysis. 
\end{abstract}



\IEEEpeerreviewmaketitle


\section{Introduction}

 \IEEEPARstart{R}{e}cent years have witnessed a growing interest towards Cyber-Physical systems (CPSs), 
 namely systems that exhibit a tight conjoining of communication, computational and physical units. 
 The fact that breaches in the cyber-space can have consequences in the physical domain has triggered considerable 
 attention towards the issue of cyber-physical security \cite{CSS-Magazine, Sastry-survey}. 
 In CPSs, attacks to the communication links can be classified as either deception attacks or Denial-of-Service (DoS) attacks. The former affect the trustworthiness of data by manipulating the packets transmitted over the network; see \cite{Tabuada}-\cite{ICM-Geometric} and the references therein. DoS attacks are instead primarily intended to affect the timeliness of the information exchange, \emph{i.e.}, to cause packet losses. This paper is concerned with DoS attacks, and,
in particular, with \emph{jamming} attacks \cite{Xu, David}, 
although in this paper we shall use these two terms interchangeably.

In the literature, the issues of securing robustness of CPSs against DoS 
has been widely investigated only for centralized architectures 
\cite{sastry}\nocite{Basar,Befekadu,Andre,Martinez,Claudio-Pietro-IFAC,Claudio-Pietro}-\cite{Claudio-Pietro-CDC14}.
On the other hand, very little is known about DoS for {distributed} coordination problems.
In this paper,
we investigate the issue of DoS with respect to consensus-like networks.
Specifically, inspired by \cite{Claudio-Paolo}, we consider a \emph{self-triggered} consensus network, in which communication and control actions are planned ahead in time, depending on the information currently available at each agent. 
The attacker objective is to prevent consensus by denying communication among the network agents.
Consensus is a prototypical problem in distributed settings with a huge range of applications, spanning from formation and cooperative robotics to surveillance and distributed computing; see for instance \cite{Claudio-Paolo}-\cite{Cortes}. 
On the other hand, self-triggered coordination turns out to be of major interest when consensus has to be achieved
in spite of possibly severe communication constraints. In this respect, a remarkable feature of self-triggered 
coordination lies in the possibility of ensuring consensus properties in the absence of any
global information on the graph topology and with no need to synchronize the agents local clocks.

A basic question in the analysis of distributed coordination in the presence of DoS 
is concerned with the modeling of DoS attacks. In \cite{Claudio-Pietro-IFAC,Claudio-Pietro}, 
a general model is considered that only constrains DoS attacks in terms of their average frequency and duration,
which makes it possible to capture many different types of DoS attacks, including trivial, periodic, random and protocol-aware jamming attacks \cite{Xu,David,XuW,Tague}. 
Building on \cite{Claudio-Pietro}, a preliminary analysis of
consensus networks in the presence of DoS is presented in \cite{Danial-Pietro-Claudio} under the 
simplifying assumption that the occurrence of DoS cause all the network links 
to fail simultaneously. This scenario is representative of networks 
operating through a single access point, in the so-called ``infrastructure" mode. 
In this paper, we consider the more general scenario in which the network communication links
can fail independent of each other, thereby extending the analysis to ``ad-hoc" (peer-to-peer) networks.
One contribution of this paper is an explicit characterization of the frequency 
and duration of DoS at the various network links under which consensus can be preserved
by suitably designing time-varying control and communication policies.
Moreover, an explicit characterization of the effects of DoS on the consensus time is provided.

Since DoS induces communication failures, the problem of achieving consensus under DoS can be naturally cast as a consensus problem for networks with 
switching topologies. This approach is certainly not new in the literature. In \cite{Murry}, for instance, it is shown that consensus can be reached whenever graph connectivity is preserved point-wise in time; \cite{Arcak} considers a notion of \emph{Persistency-of-Excitation} (PoE), which stipulates that graph connectivity should be established over a period of time, rather than point-wise in time, which is similar to the joint connectivity assumption in \cite{Morse}. 
In CPSs, however, the situation is different. In CPSs, one needs to deal with the fact that networked communication is inherently digital, which means that the rate at which the transmissions are scheduled cannot be arbitrarily large. 
Under such circumstances, the aforementioned tools turn out be ineffective. 
In order to cope with this situation, we introduce a notion of \emph{Persistency-of-Communication} (PoC),
which naturally extends the PoE condition to a digital networked setting by requiring graph (link) 
connectivity over periods of time that are consistent with the constraints imposed by the communication medium. 
A characterization of DoS frequency and duration under which consensus properties can be preserved is then obtained  by exploiting the PoC condition. 

The remainder of this paper is as follows. In Section II, we formulate the control problem 
and provide prototypical results for self-triggered consensus. In Section III, we describe 
the considered class of DoS signals.  The main results of this paper are presented in Section IV. 
In Section V, we provide a detailed discussion of the results, and show how the analysis can be extended 
so as to account for genuine (non-malicious) transmission failures. 
A numerical example is presented  in Section VI. Section VII ends the paper with concluding remarks.


\section{Self-triggered consensus network}

\subsection{System definition}

We consider a consensus network, which is represented by an undirected graph ${\mathcal G}=({\mathcal I},{\mathcal E})$, where 
${\mathcal I}=\until{n}$ denotes the node set and 
${\mathcal E} \subseteq \mathcal I \times \mathcal I$ denotes the edge set. 
Specifically, we denote by $D$ and $L$ the incidence and Laplacian matrix of 
${\mathcal G}$, respectively. For each node $i\in {\mathcal I}$, we denote by $\neigh{i}$ the set of its neighbors, and by 
$\deg^i =|\neigh{i}|$, \emph{i.e.}, the cardinality of $\neigh{i}$.
Throughout the paper, we shall refer to ${\mathcal G}$ as the ``nominal"
network, and we shall assume that ${\mathcal G}$ is connected.
 
The consensus network of interest employs \emph{self-triggered} communication 
\cite{Claudio-Paolo}, defined via
hybrid dynamics, with state variables
$(x,u,\theta)\in\real^{n}\times \real^{d} \times \real^{d}$,
where $x$ is the vector of nodes states, $u$ is the vector of controls,
$\theta$ is the vector of clock variables, and $d$
is the sum of the neighbors of all the nodes, \emph{i.e.},  
$d:=\sum_{i=1}^n d^i$. 
The control signals are assumed to belong to $\mathcal T : = \{-1,0,+1\}$. 
The specific quantizer of choice is $\map{\sign_\eps}{\real}{\mathcal T}$, which is given by
\begin{eqnarray} \label{eq:signeps}
\sign_\eps(z)
:=\begin{cases} \sign(z) & \text{if}\: |z|\ge\eps\\
0 & \text{otherwise}
\end{cases}
\end{eqnarray}
where $\eps>0$ is a sensitivity parameter, which can be used at the design 
stage for trading-off frequency of the transmissions vs. accuracy of the 
consensus region.

The system $(x,u,\theta)\in\real^{n}\times \real^{d} \times \real^{d}$ satisfies the continuous 
evolution
\begin{eqnarray} 
\label{eq:modelloA-cont-linksampling}
\arraycolsep=3pt\def\arraystretch{1.2}
\left\{ \begin{array}{l}
\dot x^i=\displaystyle \sum_{j\in\neigh{i}}u^{ij}\\
\dot u^{ij}=0\\
\dot \theta^{ij}=-1
\end{array} \right.
\end{eqnarray}
where $i\in {\mathcal I}$ and $j\in \neigh{i}$. 
The system satisfies the differential equation above for all $t$ except for those values of the time at which the set
\begin{eqnarray} 
{\cal J}(\theta,t)=\setdef{(i,j) \in {\mathcal I \times \mathcal I}}{ 
{j\in \neigh{i}} \;  \text{and} \; 
\theta^{ij}(t^-)=0}
\end{eqnarray}
is non-empty. At these times, in the ``nominal'' operating
mode (when communication between nodes is always possible),
a discrete transition occurs, which is governed by the following discrete update:
\begin{eqnarray}
\label{eq:modelloA-disc-linksampling}
\arraycolsep=3pt\def\arraystretch{1.2}
\left\{ \begin{array}{l}
x^i(t)=x^i(t^-) \quad \forall i\in {\cal I}\\
u^{ij}(t)=
\arraycolsep=3pt\def\arraystretch{1.2}
\left\{ \begin{array}{ll}
\sign_{\eps}\!\big({\cal D}^{ij}(t)\big) & \quad \text{if}\: (i,j) \in {\cal J}(\theta,t)\\
u^{ij}(t^-) & \quad \text{otherwise}
\end{array} \right. \\
 \theta^{ij}(t)=
 \arraycolsep=3pt\def\arraystretch{1.2}
\left\{ \begin{array}{ll}
f^{ij}(x(t)) & \qquad \quad \text{if}\: (i,j) \in {\cal J}(\theta,t)\\
\theta^{ij}(t^-) & \qquad \quad \text{otherwise}
\end{array} \right.
\end{array} \right.
\end{eqnarray}
where for every $i\in {\cal I}$ and $j\in \neigh{i}$, the map $\map{f^{ij}}{\real^n}{\realpositive}$ is defined by 
\begin{eqnarray}
\label{eq:linksam}
f^{ij}(x(t)) :=
 \arraycolsep=3pt\def\arraystretch{1.2}
\left\{ \begin{array}{ll}
\dst\frac{|{\cal D}^{ij}(t)|}{2 (\deg^i+\deg^j)}& \quad \text{if}\: |{\cal D}^{ij}(t)|\ge\eps\\
\dst\frac{\eps}{2(\deg^i+\deg^j)} & \quad \text{if}\: |{\cal D}^{ij}(t)|<\eps
\end{array} \right.
\end{eqnarray}
and
\begin{eqnarray}
\label{D}
{\cal D}^{ij}(t)=x^j(t)-x^i(t)
\end{eqnarray}
Notice that for all $\{i,j\} \in {\cal E}$ we have 
$\theta^{ij}(t)=\theta^{ij}(t)$ and $u^{ij}(t)=-u^{ij}(t)$ for all 
$t \in \mathbb R_{\geq 0}$. As such, the system
(\ref{eq:modelloA-cont-linksampling})-(\ref{eq:modelloA-disc-linksampling})
can be regarded as an edge-based consensus protocol.
Here, the term ``self-triggered'', first adopted in 
the context of real-time systems \cite{Velasco}, expresses the property that 
the data exchange between nodes is driven by local clocks, which avoids the need 
for a common global clock. 

 \subsection{Prototypical result for self-triggered consensus} 
 
The following result characterizes the limiting behavior 
of the system (\ref{eq:modelloA-cont-linksampling})-(\ref{eq:modelloA-disc-linksampling}).

\begin{theorem} 
\label{thm:modelloA-Consensus}
\cite{Claudio-Paolo} Let $x$ be the solution to 
(\ref{eq:modelloA-cont-linksampling})-(\ref{eq:modelloA-disc-linksampling}).
Then, for every initial condition,
$x$ converges in finite time to a point $x^* \in \mathbb R^n$ belonging to the 
set 
\begin{eqnarray}
\label{set.E}
{\mathscr E}=
\setdef{x\in \R^n}{|x^i(t)-x^j(t)|< \delta \quad \forall\, (i,j) \in {\cal I}\times {\cal I}}
\end{eqnarray}
where $\delta=\varepsilon (n-1)$.
\qedp
\end{theorem}

Theorem \ref{thm:modelloA-Consensus} will be used as a reference frame for the 
analysis of Section IV and V. This theorem is prototypical in the sense  
that it serves to illustrate the salient features of the problem of 
consensus/coordination in the presence of communication interruptions. 
Following \cite{Claudio-Paolo}, the analysis of this paper could be extended 
to include important aspects such as \emph{quantized communication}, \emph{delays}
and \emph{asymptotic consensus} (rather than practical consensus 
as in (\ref{set.E})). While important, these aspects do not add much to 
the present investigation and will be therefore omitted. 
We refer the interested reader to \cite{Claudio-Paolo} for a discussion
on how these aspects can be dealt with.


\section{Network Denial-of-Service}

We shall refer to Denial-of-Service (DoS, in short) as the phenomenon by which 
communication between the network nodes is interrupted. 
We shall consider the very general scenario
 in which the network communication links
can fail independent of each other. From the perspective of modeling,
this amounts to considering multiple DoS signals, one for each 
network communication link. 

\subsection{DoS characterization} 

Let $\{h_n^{ij}\}_{n\in \integernonnegative}$
with ${h_0^{ij}} {\ge 0} $ denote the sequence of DoS off/on transitions affecting the link $\{i,j\}$, namely the 
sequence of time instants at which the DoS status on the link $\{i,j\}$
exhibits a transition from zero (communication is possible) to one (communication is interrupted). Then
\begin{equation} \label{eq:Hn}
H_n^{ij}: = \{h_n^{ij}\} \cup \left[ h_n^{ij},h_n^{ij} + \tau _n^{ij} \right[ 
\end{equation}
represents the $n$-th DoS time-interval, 
of a length $\tau_n^{ij}\in {\realnonnegative}$, during which communication  on the link $\{i,j\}$ is not possible. 

Given ${t,\tau}\in\real_{ \ge 0}$, with $t \ge \tau$, let 
\begin{equation} \label{eq:Xi}
\Xi^{ij} (\tau ,t) :=\mathop  \bigcup \limits_{n\in{\integernonnegative}} H_n^{ij} \bigcap {[\tau,t ]}
\end{equation}
and
\begin{equation}\label{eq:Theta}
\Theta^{ij}  (\tau ,t) := [\tau,t ] \;  \backslash \; \Xi^{ij} (\tau ,t) 
\end{equation}
where $\backslash$ denotes relative complement.
In words, for each interval $[\tau,t]$, $\Xi^{ij} (\tau ,t)$ and 
$\Theta^{ij}  (\tau ,t)$ represent the sets of time instants where communication on the link $\{i,j\}$
is denied and allowed, respectively. 

The first question to be addressed is that of determining a suitable modeling framework for  DoS. Following \cite{Claudio-Pietro},
we consider a general model that only constrains DoS attacks in terms of their average frequency and duration.
Let $n^{ij}(\tau ,t)$ denote 
the number of DoS off/on transitions  on the link $\{i,j\}$ occurring on the interval $[\tau,t ]$.

\begin{assumption}[DoS frequency]\label{asum:DoSf}
For each $\{i,j\} \in {\cal E}$, there exist $\eta^{ij}  \in\real_{ \ge 1}$ and ${\tau _f^{ij}} \in\real_{ > 0}$ such that
\begin{equation}\label{eq:f}
n^{ij} (\tau ,t) \le \eta^{ij} + \frac{t - \tau }{\tau _f^{ij}}
\end{equation}
for all ${t,\tau}\in\real_{ \ge 0}$ with $t \ge \tau$. 
\qedp
\end{assumption}

\begin{assumption}[DoS duration]\label{asum:DoSd} 
For each $\{i,j\} \in {\cal E}$,
there exist $\kappa^{ij}  \in\real_{ \ge 0}$ and ${\tau _d^{ij}} \in\real_{ > 1}$ such that
\begin{equation} \label{eq:d}
\lvert \Xi^{ij} (\tau ,t) \rvert  \le \kappa^{ij}  + \frac{t - \tau }{\tau _d^{ij}}
\end{equation}
for all ${t,\tau}\in\real_{\ge 0}$ with $t \ge \tau$.
\qedp
\end{assumption}

In Assumption 1, the term ``frequency" stems from the fact that 
$\tau _f^{ij}$ provides a measure of the ``dwell-time" between any two consecutive 
DoS intervals on the link $\{i,j\}$. The quantity $\eta^{ij}$ is needed to render (\ref{eq:f}) 
self-consistent when $t=\tau=h_n^{ij}$ for some $n \in \mathbb Z_{\geq 0}$, in which case
$n^{ij} (\tau ,t)=1$. Likewise, in Assumption 2, the term ``duration"
is motivated by the fact that $\tau _d^{ij}$ provides a measure of the fraction of  
time (${\tau _d^{ij}} > 1$) the link $\{i,j\}$ is under DoS. 
Like $\eta^{ij}$, the constant $\kappa^{ij}$ plays the role
of a regularization term. It is needed because
during a DoS interval, one has $|\Xi(h_n^{ij},h_n^{ij}+\tau_n^{ij})| = \tau_n^{ij} \geq  \tau_n^{ij} /\tau _d^{ij}$
since ${\tau _d^{ij}}>1$, with $ \tau_n^{ij} =  \tau_n^{ij} /\tau _d^{ij}$ if and only if  $ \tau_n^{ij} =0$.
Hence, $\kappa^{ij}$ serves to make (\ref{eq:d}) self-consistent.
Thanks to the quantities $\eta^{ij}$ and $\kappa^{ij}$, DoS frequency and duration are both average quantities. 

\begin{remark}
Throughout this paper, we will mostly focus on the case where DoS
is caused by malicious attacks. Of course, DoS might also result from a
``genuine" network congestion. 
We shall briefly address this case in Section V-C.
\qedp
\end{remark}


\subsection{Examples}
The considered assumptions only pose 
limitations on the frequency of the DoS status and its duration. As such, 
this characterization can capture many different 
scenarios, including trivial, periodic, random and protocol-aware jamming attacks \cite{Xu,David,XuW,Tague}. For the sake of simplicity,
we limit out discussion to the case of radio frequency (RF) jammers, although 
similar considerations can be made with respect to spoofing-like threats \cite{Bellardo}.

Consider for instance the case of
\emph{constant jamming}, which is one of the most common threats that may occur in a wireless network
\cite{Xu,Pelechrinis}. By continuously emitting RF signals on the wireless medium, 
this type of jamming can lower the Packet Send Ratio (PSR)
for transmitters employing carrier sensing as medium access policy 
as well as lower the Packet Delivery Ratio (PDR) by corrupting packets at the receiver. 
In general, the percentage of packet losses caused by this type of jammer
depends on the Jamming-to-Signal Ratio and can be difficult to quantify 
as it depends, among many things, on the type
of anti-jamming devices, the possibility to adapt the signal strength threshold
for carrier sensing, and the interference signal power, which may vary with time. 
In fact, there are several provisions that can be taken
in order to \emph{mitigate} DoS attacks, including spreading techniques, high-pass filtering
and encoding \cite{DeBruhl,Tague}. These provisions 
decrease the chance that a DoS attack will be successful,
and, as such, limit in practice the frequency and duration of the time intervals
over which communication is effectively denied. This is nicely 
captured by the considered formulation. 

As another example, consider the case of \emph{reactive jamming} \cite{Xu,Pelechrinis}.
By exploiting the knowledge of the 802.1i MAC layer protocols, a jammer may  
restrict the RF signal to the packet transmissions. The collision period need not be long since 
with many CRC error checks a single bit error can corrupt an entire frame.
Accordingly, jamming takes the form of a (high-power) burst of noise,
whose duration is determined by the length of the symbols to corrupt \cite{DeBruhl,Wood}. 
Also this case can be nicely accounted for via the considered assumptions.


\section{DoS-resilient consensus}
 

\subsection{Modified communication protocol}


In order to achieve robustness against DoS, the nominal discrete evolution  
\eqref{eq:modelloA-disc-linksampling} is modified as follows:

{
\small 
{\setlength\arraycolsep{-1pt} 
\begin{eqnarray}
\label{eq:modelloA-disc-linksampling-DoS}
 \left\{
 \arraycolsep=3pt\def\arraystretch{1.5}
\begin{array}{ll}
x^i(t)=x^i(t^-) \quad \forall i\in {\cal I} \\
u^{ij}(t)=
\left\{
\arraycolsep=3pt\def\arraystretch{1.5}
\begin{array}{ll}
\sign_{\eps}\!\left({\cal D}^{ij}(t)\right) & \text{if}\: (i,j)\in {\cal J}(\theta,t) \wedge t\in \Theta^{ij}(0,t) \\
0 & \text{if}\: (i,j)\in {\cal J}(\theta,t) \wedge t\in \Xi^{ij}(0,t)  \\
u^{ij}(t^-)  &  \text{otherwise}
\end{array} \right. \\
 \theta^{ij}(t)=
 \left\{
 \arraycolsep=3pt\def\arraystretch{1.5}
\begin{array}{ll}
f^{ij}(x(t)) & \quad \,  \text{if}\: (i,j)\in {\cal J}(\theta,t) \wedge t\in \Theta^{ij}(0,t) \\
\displaystyle \frac{\eps}{2(\deg^i+\deg^j)} & \quad \, \text{if}\: (i,j)\in {\cal J}(\theta,t) \wedge t\in \Xi^{ij}(0,t)  \\
\theta^{ij}(t^-)  & \quad \,  \text{otherwise}
\end{array} \right. 
\end{array} \right. \nonumber \\
\end{eqnarray}
}}%
\normalsize
In words, the control action $u^{ij}$ is reset to zero whenever 
the link $\{i,j\}$ is in DoS status. Notice that this requires that 
the nodes are able to detect the occurrence of DoS. This is the case, 
for instance, with transmitters employing carrier sensing as medium access policy.
Under such circumstances, a DoS signal in the form of \emph{constant jamming}
(\emph{cf.} Section III-B) can be detected. 
Another example is when transceivers use TCP acknowledgment
and DoS takes the form of \emph{reactive jamming} (\emph{cf.} Section III-B). 
In addition to $u$, also the local clocks are modified upon DoS, yielding
a \emph{two-mode} sampling logic. 
In particular,  for each $\{i,j\}\in {\cal E}$, let
$\{t^{ij}_k\}_{k\in \integernonnegative}$ denote  the sequence of transmission attempts.
Then, each $\theta^{ij}$ satisfies
\begin{eqnarray}
\label{eq:st}
t_{k+1}^{ij}=t_k^{ij} +\left\{ \begin{array}{ll}
{f^{ij}}(x(t_k^{ij})) & \quad \,  \text{if}\:  t_k^{ij} \in \Theta^{ij}(0,t)\\ \\
\displaystyle \frac{\eps}{2(\deg^i+\deg^j)} & \quad \,   \text{otherwise}
\end{array} \right. 
\end{eqnarray}
As it will become clear later on, this is in
order to maximize the robustness of the consensus protocol
against DoS. By (\ref{eq:st}), it is an easy matter to see that for each $\{i,j\}\in {\cal E}$
 the sequences $\{t^{ij}_k\}_{k\in \integernonnegative}$
satisfy a ``dwell-time'' property, since 
\be\label{eq:DeltaT-bound-Nodesampling}
\Delta_{k}^{ij}:=t^{ij}_{k+1}-t^{ij}_{k} \, \ge \, \frac{\eps}{4 \degmax}
\ee   
for all $k\in \mathbb R_{\geq 0}$, where $d_{max}=\max_{i \in {\cal I}} d^i$.
This ensures that all the sequences of transmission times are Zeno-free.

For the sake of clarity, the DoS-resilient consensus protocol is summarized below.

\smallskip
\myrule
\vspace{.75\smallskipamount}
\noindent\hfill\textbf{DoS-resilient consensus protocol  
\stopmodif}\hfill\vspace{.75\smallskipamount}
\myrule\vspace{.75\smallskipamount}
\begin{algorithmic}[1]
\STATE {\bf initialization:} 
For all $i\in \mathcal I$ and $j\in \mathcal N_i$, set $\theta^{ij}(0^-)=0$, $u^{ij}(0^-)\in \{-1,0,+1\}$, 
and $u^i(0^-)=\sum_{j\in \mathcal N_i} u^{ij}(0^-)$;
\FORALL{$i\in \mathcal I$}
\FORALL{$j\in \mathcal N_i$}
\WHILE{$\theta^{ij}(t)>0$} 
\STATE $i$ applies the control $u^i(t)=\sum_{j\in \mathcal N_i} u^{ij}(t)$;
\ENDWHILE
\IF{$\theta^{ij}(t^-)=0 \wedge t\in \Theta^{ij}(0,t)$}
\STATE $i$ updates $u^{ij}(t)=\sign_{\eps}\!\big(x^j(t)-x^i(t)\big)$;
\STATE $i$ updates $\theta^{ij}(t)=f^{ij}(x(t))$;
\ELSE \IF {$\theta^{ij}(t^-)=0 \wedge t\in  \Xi^{ij}(0,t)$}
\STATE $i$ updates $u^{ij}(t)=0$;
\STATE $i$ updates $\theta^{ij}(t)=\displaystyle \frac{\eps}{2(\deg^i+\deg^j)}$;
\ENDIF
\ENDIF
\ENDFOR
\ENDFOR
\end{algorithmic}
\vspace{.5\smallskipamount}\myrule\smallskip
\medskip

\subsection{Convergence of the solutions and $\delta$-consensus}

We are now in position to characterize the overall network behavior 
in the presence of DoS. In this respect, the analysis is subdivided 
into two main steps: i) we first prove that 
all the network nodes eventually stop to update their 
local controls; and ii) we then provide conditions on the DoS frequency 
and duration such that consensus, in the sense of \eqref{set.E}, is preserved.
The latter property is achieved by resorting to a notion of
\emph{Persistency-of-Communication}, which 
determines the amount of DoS (frequency and duration) under 
which consensus can be preserved. 

As for i), the following result holds true. 

\begin{proposition} \label{prop:modelloA-convergence-linksampling}
(\emph{Convergence of the solutions})
Let $x$ be the solution to \eqref{eq:modelloA-cont-linksampling}  and \eqref{eq:modelloA-disc-linksampling-DoS}. 
Then, for every initial condition, there exists 
a finite time $T_*$ such that,for any $i \in {\cal I}$, it holds that $u^{i}(t)=0$ for all $t \geq T_*$. 
\end{proposition} 

\emph{Proof.}
Consider the Lyapunov function
\begin{eqnarray}
V(x)=\frac{1}{2}{x^\top x}
\end{eqnarray}
Let $t_k^{ij}:= \max \{t_\ell^{ij}: t_\ell^{ij} \leq t, \ell \in \mathbb Z_{\geq 0}\}$.
First notice that the derivative of $V$ along 
the solutions to (\ref{eq:modelloA-cont-linksampling}) satisfies
\be\label{eq:dotV-bound-GDoS}
\begin{aligned}
\dot V(x(t)) & = \sum_{i=1}^n
x^i(t) \dot x^i(t) \\ & = \sum_{i=1}^n  [  x^i(t)\sum_{j\in {\cal N}_i}u^{ij}(t) ] \\ 
& =-\sum_{\substack{\{i,j\}\in \E: \\ |{\cal D}^{ij}(t_k^{ij})|\ge \eps \; \wedge \; t_k^{ij} \in  \Theta^{ij}(0,t)}} 
{\cal D}^{ij}(t) \sign_{\eps} ({\cal D}^{ij}(t_k^{ij})) \\
&\le - \sum_{\substack{\{i,j\}\in \E: \\ |{\cal D}^{ij}(t_k^{ij})|\ge \eps \; \wedge \; t_k^{ij} \in  \Theta^{ij}(0,t)}} \frac{\lvert {\cal D}^{ij} (t_k^{ij}) \rvert}{2}
\end{aligned}
\ee
In words, the derivative of $V$ decreases whenever, for some $\{i,j\}\in \E$, 
two conditions are met: i) $|{\cal D}^{ij} (t_k^{ij})| \geq \eps$, which means 
that $i$ and $j$ are not $\eps$-close; and ii) 
communication on the link that connects $i$ and $j$ is possible. 
The third equality follows from the fact that for any  $\{i,j\} \in \E$
for which $|{\cal D}^{ij} (t_k^{ij})| < \eps$ or  $t_k^{ij} \in \Xi^{ij}(0,t)$ we have $u^{ij}(t)=0$
for all $[t_k^{ij},t_{k+1}^{ij}[$,
and the fact that $u^{ij}(t)= \sign_{\eps} ({\cal D}^{ij}(t_k^{ij}))$ where
${\cal D}^{ij}(t)= x^j(t)-x^i(t)$. The inequality follows from the fact that,
during the continuous evolution $|{\dot {\cal D}}^{ij}(t)| \leq d^i + d^j$
and at the jumps ${\cal D}^{ij}(t)$ does not change its value. This
implies that ${\cal D}^{ij}(t)$ cannot differ from ${\cal D}^{ij}(t_k^{ij})$
in absolute value for more than $(d^i + d^j)(t-t_k^{ij})$. Exploiting
this fact, if communication is allowed and $|{\cal D}^{ij} (t_k^{ij})| \geq \eps$
then by (\ref{eq:linksam}) and (\ref{eq:st}) we have 
\begin{eqnarray}
|{\cal D}^{ij} (t)|  \geq |{\cal D}^{ij}(t_k^{ij})|/2
\end{eqnarray}
and 
\begin{eqnarray}
\sign_{\eps} ({\cal D}^{ij}(t))=\sign_{\eps} ({\cal D}^{ij}(t_k^{ij}))
\end{eqnarray} 
for all $[t_k^{ij},t_{k+1}^{ij}[$.

From (\ref{eq:dotV-bound-GDoS}) there must exist 
a finite time $T_*$ such that, for every $\{i,j\} \in \E$ and every $k$ 
with $t_k^{ij} \geq T_*$, it holds that $\lvert {\cal D}^{ij}(t_k^{ij})\rvert < \varepsilon$ or $t_k^{ij} \in \Xi^{ij}(0,t)$.
This is because, otherwise, $V$ would become negative.
The proof follows recalling 
that in both the cases $\lvert {\cal D}^{ij}(t_k^{ij})\rvert < \varepsilon$ and 
$t_k^{ij} \in \Xi^{ij}(0,t)$
the control $u^{ij}(t)$ is set equal to zero. \qedp

The above result does not allow one to conclude 
anything about the final disagreement vector in the sense
that given a pair of nodes $(i,j)$ the asymptotic value 
of $\lvert x^j(t)-x^i(t) \rvert$ can be arbitrarily large. 
As an example, if node $i$ is  never allowed to communicate then $x^i(t)=x^i(0)$
for all $t \in \mathbb R_{\geq 0}$.
In order to recover 
the same conclusions as in Theorem 1,  
bounds on DoS frequency and duration have to be enforced.
The result which follows provides one such characterization.

Let $\{i,j\} \in {\cal E}$ be a generic network link,
and consider a DoS sequence on $\{i,j\}$,
which satisfies Assumption 1 and 2. 
Define
\begin{eqnarray}\label{eq:alpha}
\alpha^{ij} := \frac{1}{\tau^{ij}_d}+\frac{\Delta^{ij}_*}{\tau^{ij}_f} 
\end{eqnarray}   
where
\begin{eqnarray}\label{eq:Deltaij}
\Delta^{ij}_* := \frac{\varepsilon}{2(d^i+d^j)}
\end{eqnarray}   
\begin{proposition}[Link Persistency-of-Communication (PoC)] 
Consider any link $\{i,j\} \in {\cal E}$ employing the transmission 
protocol \eqref{eq:modelloA-disc-linksampling-DoS}. Also
consider any DoS sequence on $\{i,j\}$, which satisfies Assumption 1 and 2 
with $\eta^{ij}$ and $\kappa^{ij}$ arbitrary, and $\tau^{ij}_d$ and $\tau^{ij}_f$ such that
$\alpha^{ij} < 1$. Let
\begin{eqnarray}\label{eq:DoScons}
\Phi^{ij} := \frac{\kappa^{ij}+ (\eta^{ij} +1)\Delta_*^{ij}}{1-\alpha^{ij}}
\end{eqnarray}   
Then, for any given {unsuccessful} transmission 
attempt $t^{ij}_k$, at least one successful transmission occurs over the link $\{i,j\}$
within the interval $[t^{ij}_k,t^{ij}_k+\Phi^{ij}]$. 
\label{prop:persist-synchDoS} 
\end{proposition}

\emph{Proof.} In order to maintain continuity, a proof of this result is reported 
in Appendix. \qedp

We refer to the property above as a PoC condition since this property guarantees 
that DoS does not permanently destroy communication. 
Combining Proposition 1 and 2, the main result of this section
can be stated. 

\begin{theorem}[$\delta$-consensus]
Let $x$ be the solution to \eqref{eq:modelloA-cont-linksampling} and \eqref{eq:modelloA-disc-linksampling-DoS}. 
For each $\{i,j\} \in \E$, consider any DoS sequence that satisfies Assumption 1 and 2 
with $\eta^{ij}$ and $\kappa^{ij}$ arbitrary, and $\tau^{ij}_d$ and $\tau^{ij}_f$ such that
$\alpha^{ij} < 1$. Then,
for every initial condition, $x$ converges in finite 
time to a point $x^*$ belonging to the set ${\mathscr E}$ 
as in (\ref{set.E}). 
\end{theorem}

\emph{Proof.}
By Proposition 1, 
all the local controls become zero in a finite time $T_*$.
In turns, Proposition 2 excludes that this is due to the persistence of a DoS status. 
This means that, for all $\{i,j\} \in \E$,  $|{\cal D}^{ij}(t)| = \lvert x^j(t)-x^i(t) \rvert <\varepsilon$
for all $t \geq T_*$.
Since each pair of neighboring nodes differs by a most $\varepsilon$ and the nominal graph is connected, 
we conclude that each pair of network nodes can differ by at most 
$\delta=\varepsilon (n-1)$.\qedp

\subsection{Convergence time}

The above theorem shows that convergence is reached in a finite
time. The following result characterizes the effect of DoS 
on the convergence time.
\begin{lemma}[Bound on the convergence time] \label{prop:consensus-Time-synchDoS} 
Consider the same assumptions as in Theorem 1. Then,  
\begin{eqnarray} \label{eq:Consensus-Time}
T_* \leq \left[ \frac{1}{\varepsilon} + \frac{ \degmax}{\varepsilon \degmin} 
+  \frac{4 \degmax}{\varepsilon^2} \Phi  \right] \sum_{i \in \mathcal I}(x^i(0))^2
\end{eqnarray}
where $\degmin := \min_{i \in \mathcal I}d^i$ and $\Phi := \max_{\{i,j\} \in \E} \Phi^{ij}$.
\end{lemma}

\emph{Proof.}
Consider the same Lyapunov function $V$ as in the proof 
of Proposition 1. Notice that, by construction of the control law 
and the scheduling policy, for every successful transmission $t^{ij}_k$
characterized by $|{\cal D}^{ij}(t^{ij}_k)| \geq \varepsilon$, the function $V$
decreases with rate not less than $\varepsilon/2$ for at least $\varepsilon/(4 \degmax)$
units of time. Hence, $V$ decreases by a least $\varepsilon^2 / (8 \degmax) := \varepsilon_*$.
Considering all the network links, such transmissions are in total 
no more than $\lfloor V(0)/\varepsilon_* \rfloor$ since, otherwise, 
the function $V$ would become negative. Hence, it only remains 
to compute the time needed to have $\lfloor V(0)/\varepsilon_* \rfloor$ 
of such transmissions. In this respect, pick any $t_* \geq 0$
such that consensus has still not be reached. 
Note that we can have $u^{ij}(t_*)=0$ for all $\{i,j\} \in \E$.
However, this condition can last only for a limited amount of time.
In fact, if $u^{ij}(t_*)=0$ then the next transmission attempt, say $\ell^{ij}$,
over the link  $\{i,j\}$ will necessarily occur at a time 
less than or equal to $t_*+\Delta^{ij}_*$ with $\Delta^{ij}_* \leq \varepsilon/(4 \degmin)$.
Let $\mathcal Q := [t_*,t_*+ \varepsilon/(4 \degmin)]$, and suppose that
over $\mathcal Q$ all the controls $u^{ij}$ have remained equal to zero.
This implies that for some $\{i,j\} \in \E$ we necessarily have 
that $\ell^{ij}$ is unsuccessful. This is because if $u^{ij}(t)=0$ for all $\{i,j\} \in \E$
and all $t \in \mathcal Q$ then $x^{i}(t)=x^{i}(t_*)$ for all $i \in \mathcal I$
and all $t \in \mathcal Q$. Hence, if all the $\ell^{ij}$ were successful,  
we should also have $u^{ij}(\ell^{ij}) \neq 0$ for some $\{i,j\} \in \E$
since, by hypothesis, consensus is not reached at time $t_*$.
Hence, applying Proposition 2 we conclude that at least one of
the controls $u^{ij}$ will become non zero before $\ell^{ij}+\Phi^{ij}$ 
units of time have elapsed. Overall, this implies that at least one 
control will become nonzero before $\varepsilon/(4 \degmin) + \Phi$ 
units of time have elapsed. Since $t_*$ is generic,
we conclude that $V$ decreases by at least $\varepsilon_*$ 
every  $\varepsilon/(4 \degmax) + \varepsilon/(4 \degmin) + \Phi$ 
units of time, which implies that
\begin{eqnarray} 
T_* \leq 
\left[ \frac{\varepsilon}{4 \degmax} + \frac{\varepsilon}{4 \degmin} + \Phi  \right] 
\frac{V(0)}{\varepsilon_*}  
\end{eqnarray} 
The thesis follows by recalling that $V(0)$ can be rewritten 
as $V(0)=\frac{1}{2} \sum_{i \in \mathcal I}(x^i(0))^2 $. \qedp

\section{Discussion and extensions}

\subsection{Persistency-of-Communication and consensus under
permanent link disconnections}

%

As it follows from the foregoing analysis, consensus is achieved 
whenever for each link $\{i,j\} \in \E$, the DoS signal satisfies 
$\alpha^{ij} < 1$. This condition poses limitations on 
both DoS frequency and duration. It is worth noting that this condition
is in a wide sense also necessary in order to achieve consensus. 
To see this, consider a network for which removing the link $\{i,j\}$
causes the network underlying graph to be disconnected. 
Of course, if communication over $\{i,j\}$ is always denied then
consensus cannot be achieved for arbitrary initial conditions.
In this respect, it is an easy matter to see that condition $\alpha^{ij} < 1$
becomes necessary to achieve consensus. In fact,
denote by $\mathcal S(\tau^{ij}_f,\tau^{ij}_d)$ the class of all DoS
signals for which $\alpha^{ij} \geq1$. Then,
$\mathcal S(\tau^{ij}_f,\tau^{ij}_d)$ does always contain DoS signals for which 
communication over the link $\{i,j\}$ can be permanently denied.
As an example, consider the DoS signal characterized by 
$(h^{ij}_n,\tau^{ij}_n)=(t^{ij}_k,0)$. This DoS signal satisfies 
Assumption 1 and 2 with $(\eta^{ij},\kappa^{ij},\tau^{ij}_f,\tau^{ij}_d) =(1,0,\Delta^{ij}_*,\infty)$, 
but destroys any communication attempt
over the link $\{i,j\}$.
As another example, consider the DoS signal characterized by
$(h^{ij}_0,\tau^{ij}_0)=(0,\infty)$.
This signal satisfies Assumption 1 and 2 with 
$(\eta^{ij},\kappa^{ij},\tau^{ij}_f,\tau^{ij}_d) =(1,0,\infty,1)$,
but, as before, destroys any communication attempt
over the link $\{i,j\}$. In both the examples, $\alpha^{ij} = 1$. 

Requiring $\alpha^{ij} < 1$ is not surprising.
In fact, the fulfillment of this condition requires
that 
\begin{eqnarray} \label{eq:constraints}
{\tau^{ij}_f} > \Delta^{ij}_* \quad \text{and} \quad \tau^{ij}_d > 1
\end{eqnarray}
The first requirement, ${\tau^{ij}_f} > \Delta^{ij}_*$,
simply means that DoS can occasionally occur at a rate 
faster than the highest transmission rate of the link $\{i,j\}$.
However, on the average, the frequency at which DoS can occur 
must be sufficiently small compared to sampling rate of the network link.  
Likewise, the second requirement, ${\tau^{ij}_d} > 1$, simply 
means that, on the average, the amount of DoS affecting link $\{i,j\}$
must necessarily be a fraction of the total time.  
PoC can be therefore regarded as an average connectivity property. 

It is worth noting that in some cases consensus can be preserved 
even if $\alpha^{ij} \geq 1$ for certain network links. 
This happens whenever removing such links does not 
cause the graph to de disconnected.  
More precisely, let $\mathcal X$ be any set of links 
such that $\mathcal G_{\mathcal X} : = (\mathcal I, \mathcal E \setminus \mathcal X)$
remains connected. From the foregoing analysis, it is immediate to conclude that
consensus is preserved whenever $\alpha^{ij} < 1$ 
for all $\{i,j\} \in \mathcal E \setminus \mathcal X$, 
even if communication over the links $\{i,j\} \in \mathcal X$
is permanently denied.

\subsection{Comparison with classic connectivity conditions}

As previously noted, PoC
can be regarded as an average connectivity property as it does not 
require graph connectivity point-wise in time.
In this sense, it is reminiscent of 
\emph{Persistency-of-Excitation} conditions that are found in 
the literature on consensus under switching topologies (\emph{e.g.}, see \cite{Arcak}).
There are, however, noticeable differences. To see this, consider 
the simple situation in which the Dos pattern is the same for all the links, 
\emph{i.e.}, $(h^{ij}_n,\tau^{ij}_n)=(h_n,\tau_n)$ for all $\{i,j\} \in \E$ and 
all $n \in \mathbb Z_{\geq 0}$.
Under such circumstances, the incidence matrix of the graph is a time-varying 
matrix satisfying: i) $D(t)=0$ in the presence of DoS; and ii)
$D(t)=D$ in the absence of DoS, where $D$ represents 
the incidence matrix related to the nominal graph configuration.
Consider now a DoS pattern consisting of countable number of singletons, 
\emph{i.e.}, $H_n = \{h_n\}$ for all $n \in \mathbb Z_{\geq 0}$.
In a classic continuous-time setting, such a DoS pattern does not 
destroy consensus.
In fact, it is trivial to conclude that there exist 
constants $c_1,c_2 \in \real_{> 0}$
such that (\emph{cf}.  \cite{Arcak})
\begin{eqnarray}
\int_{t_0}^{t_0+c_1} Q D(t)D^\top(t) Q^\top dt= Q DD^\top Q^\top c_1 > c_2 I
\end{eqnarray}
for all $t_0 \in \real_{\ge 0}$, where $Q$ is a suitable projection matrix
such that $Q D(t)D^\top(t) Q^\top$ is nonsingular if and only if 
the graph induced by $D(t)$ is connected.
In the present case, in accordance 
with the previous discussion, consensus can instead be destroyed.
The subtle, yet important, difference is due to the constraint 
on the frequency of the information exchange that is 
imposed by the network.
In this sense, the notion of PoC naturally extends the 
{Persistency-of-Excitation} condition 
to digital networked settings by requiring that the graph connectivity
be established over periods of time that are consistent with the maximum 
transmission rate imposed by the communication protocol. 

\subsection{Accounting for genuine DoS}

In the foregoing analysis, we focused on the case where DoS
is caused by malicious attacks. Of course, DoS might also result from a
``genuine" network congestion. Hereafter, we will briefly discuss
how the case of genuine DoS can be incorporated into the present 
framework. We shall focus on a deterministic formulation of the problem.
A probabilistic characterization of the problem, though restricted to a centralized 
setting, has been proposed in \cite{Ishii}.

Let $\beta^{ij} \in [0,1]$ be an upperbound on the average
percentage of transmission failures that can occur over the link $\{i,j\}$. 
This bound can be chosen as representative of the situation where 
all the network nodes exchange information at the highest transmission rate 
(according to (\ref{eq:st}), this is equal to $4\degmax/\varepsilon$ for each link).
Here. by ``average" we mean that, denoting by $T^{ij}_A(\tau,t)$ and 
$T^{ij}_F(\tau,t)$ the number of transmission attempts and transmission failures 
for the link $\{i,j\}$ on the interval $[\tau,t]$, it holds that 
\begin{eqnarray} \label{eq:genuineDoS}
\frac{T^{ij}_F(\tau,t)}{T^{ij}_A(\tau,t)} \leq \beta^{ij} 
\end{eqnarray}
as $T^{ij}_A(\tau,t) \rightarrow \infty$.

This condition can be suitably rearranged.
To this end, first notice that the above condition is equivalent to
the existence of a positive constant $a^{ij}$ such that 
\begin{eqnarray} \label{eq:genuineDoS_equiv}
T^{ij}_F(\tau,t) \leq a^{ij} + \beta^{ij} T^{ij}_A(\tau,t)   
\end{eqnarray}
for all $t,\tau \in \mathbb R_{\geq 0}$ with $t \geq \tau$.
Moreover, it holds that
$
T^{ij}_A(\tau,t)  \leq  \lceil (t - \tau)/\Delta_*^{ij} \rceil
$
since, by construction, $\Delta_*^{ij}$ is the smallest inter-transmission  
time for the link $\{i,j\}$. Letting $b^{ij}:=a^{ij} +1$, we then have
\begin{eqnarray}
T^{ij}_F(\tau,t) \leq b^{ij} + \frac{t-\tau}{(\Delta_*^{ij} /  \beta^{ij})}
\end{eqnarray}
Therefore, we can regard genuine transmission failures as the result 
of a DoS signal in the form of a train of pulses that are superimposed to the 
transmission instants, where $T^{ij}_F(\tau,t)$ coincides with the number $n^{ij}(\tau,t)$
of DoS off/on transitions occurring on the interval $[\tau,t]$.
Thus, Assumption 1 and 2 are satisfied with 
$(\eta^{ij},\kappa^{ij},\tau^{ij}_f,\tau^{ij}_d) =(b^{ij},0,\Delta_*^{ij} /  \beta^{ij},\infty)$.
According to the analysis of Section IV, one can conclude the following:
i) if only genuine transmission failures are present (no malicious DoS),
Persistency-of-Communication is preserved as long as
\begin{eqnarray}
\frac{1}{\tau^{ij}_d}+\frac{\Delta^{ij}_*}{\tau^{ij}_f} = \beta^{ij}  < 1
\end{eqnarray}  
This is consistent with intuition and, in fact, simply means that communication 
over the link $\{i,j\}$ is not permanently destroyed if and only if
$T^{ij}_F(\tau,t) < T^{ij}_A(\tau,t)$ on the average;
ii) in case of genuine and malicious transmission failures,
one can simply consider two independent DoS signals acting on the same link,
each one characterized by its own 4-tuple $(\eta^{ij},\kappa^{ij},\tau^{ij}_f,\tau^{ij}_d)$.
It is immediate to see that that the analysis of Section IV carries over to 
the present case by replacing condition $\alpha^{ij} < 1$ with 
$\alpha^{ij} + \beta^{ij} < 1$. 


\section{A numerical example}
We consider a random connected undirected graph with $n=40$ 
nodes and with $d^i=4$ for all $i \in \mathcal I$. 
Nodes and control initial values are generated randomly within 
the interval $[0,1]$ and the set $\{-1,0,1\}$, respectively.

\begin{figure}[t]
	\centering
		\includegraphics[width=1\columnwidth]{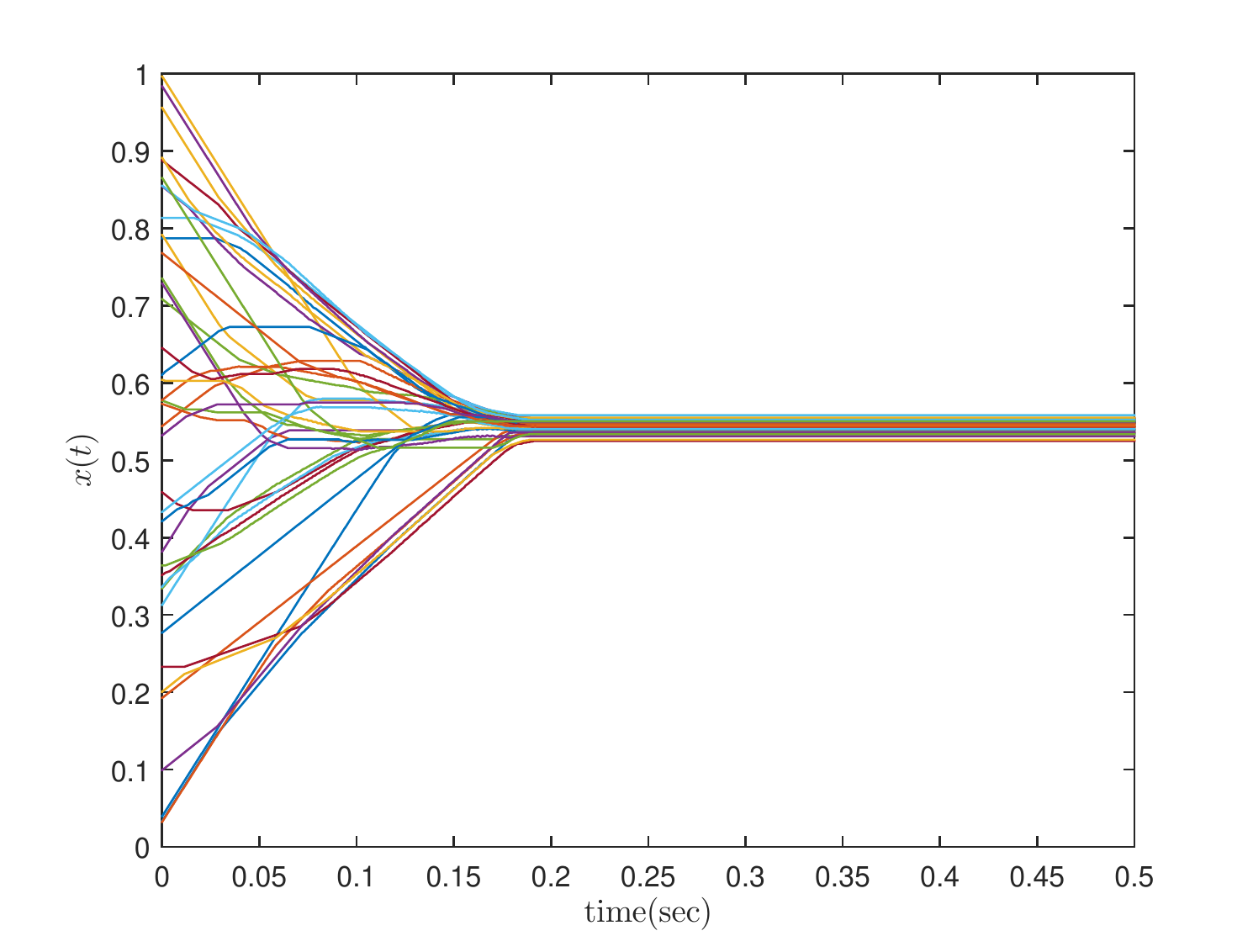}
	\caption{Evolution of $x$, corresponding to the solution to \eqref{eq:modelloA-cont-linksampling} and \eqref{eq:modelloA-disc-linksampling-DoS} for a random graph with $n=40$ nodes 
	in the absence of DoS.}
	\label{fig:x}
\end{figure}

We consider the behavior of \eqref{eq:modelloA-cont-linksampling} 
and \eqref{eq:modelloA-disc-linksampling-DoS} with $\varepsilon=0.005$.
Figure 1 depicts simulation results for the nominal case in which DoS 
is absent. Notice that in this case \eqref{eq:modelloA-disc-linksampling-DoS} coincides 
with \eqref{eq:modelloA-disc-linksampling}.
We next consider the case in which DoS is present.
Simulation results are reported in Figure 2. In the simulation,
we considered DoS attacks which affect each of the network links independently.
For each link, the corresponding DoS pattern takes the form of a pulse-width modulated signal
with variable period and duty cycle
(maximum period of $0.15$sec and maximum duty cycle equal to $100\%$), both generated randomly. 
These patterns are reported in  Table I and depicted in Figure 3 
for a few number of network links.
Notice that, for each DoS pattern, one can compute corresponding values for 
$(\eta^{ij},\kappa^{ij},\tau_f^{ij},\tau_d^{ij})$. 
They can be determined by computing the values  $n^{ij} (\tau ,t)$ and $|\Xi^{ij} (\tau,t)|$ 
of each DoS pattern (\emph{cf.} Assumption 1 and 2) over the considered simulation horizon.
Figure 4 depicts the obtained values of $\tau _f^{ij}$ and $\tau _d^{ij}$
for each $\{i,j\} \in \E$. One sees that these values are consistent 
with the requirements imposed by the PoC condition.

\begin{figure}
	\centering
		\includegraphics[width=1\columnwidth]{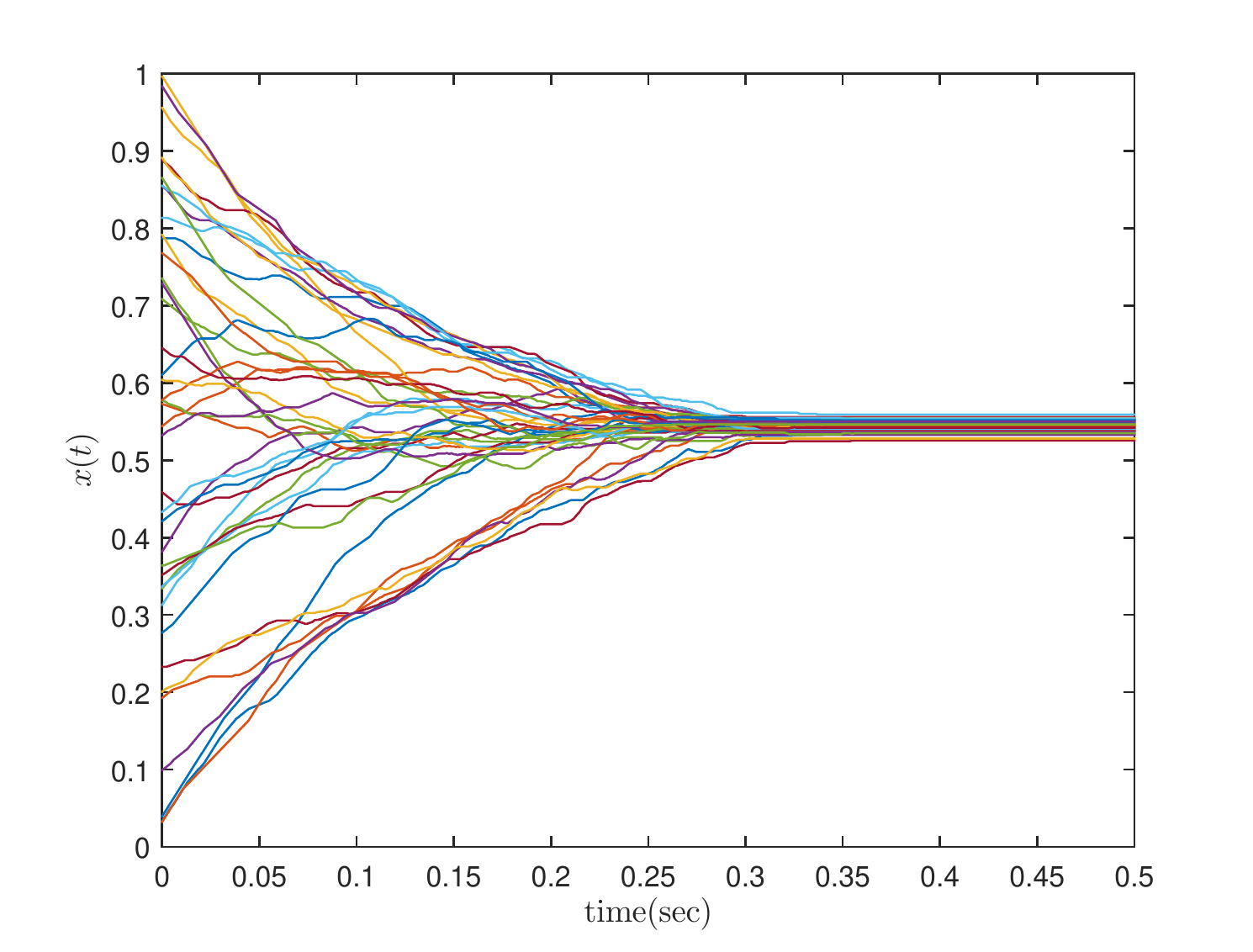}
	\caption{Evolution of $x$, corresponding to the solution to \eqref{eq:modelloA-cont-linksampling} and \eqref{eq:modelloA-disc-linksampling-DoS} for a random graph with $n=40$ nodes 
	in the presence of DoS.}
	\label{fig:x2}
\end{figure}


\begin{figure}
	\centering
		\includegraphics[width=1\columnwidth]{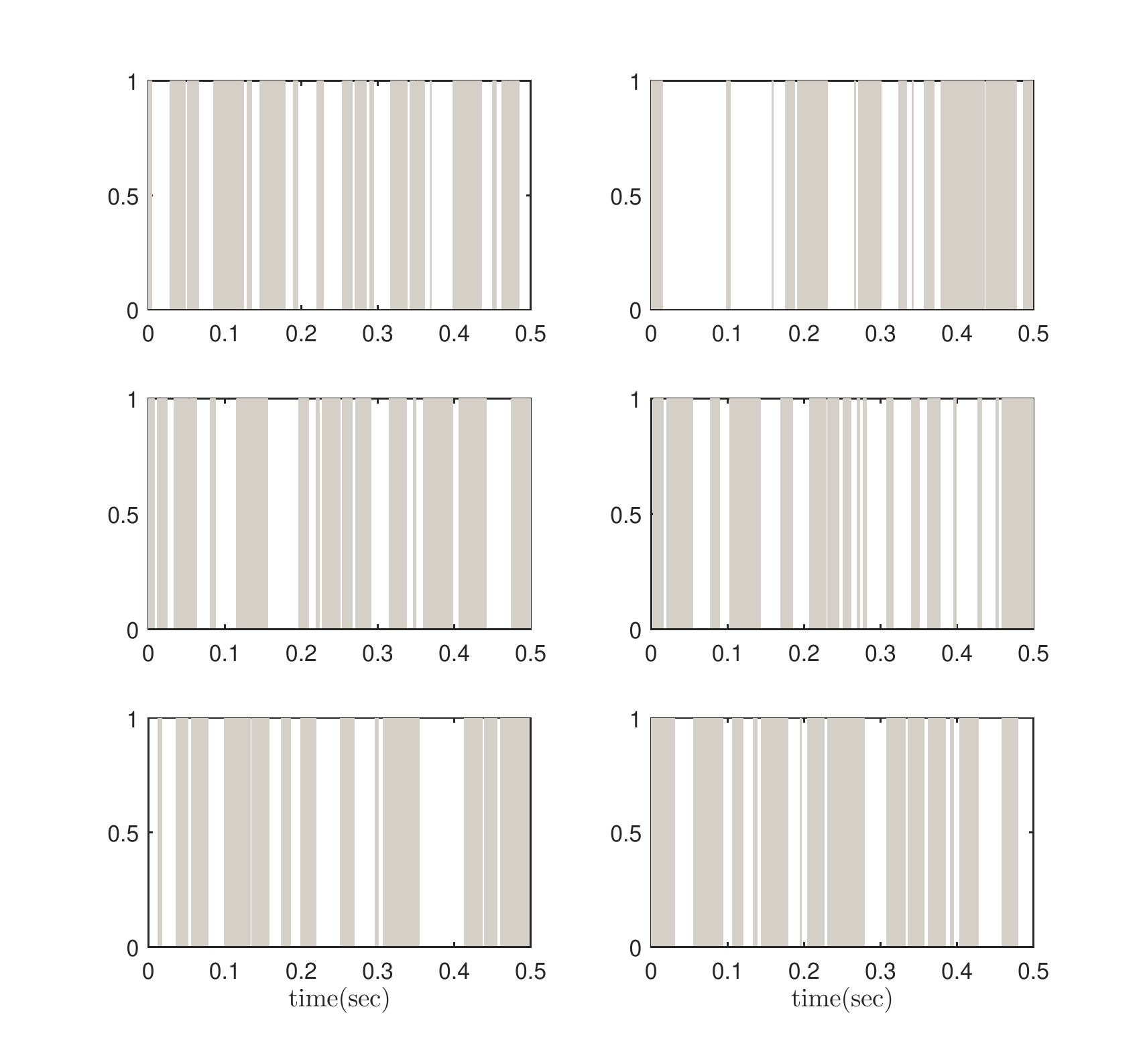}
	\caption{DoS pattern for the network links $\{13,14\}$, $\{6,34\}$, $\{34,39\}$, $\{9,26\}$, $\{9,21\}$ and $\{33,38\}$.
	The vertical gray stripes represent the time-intervals over which DoS is active.}
	\label{fig:x3}
\end{figure}

\begin{table}%
\centering
\caption{DoS average duty cycle over some links }
\begin{tabular}{|c | c || c | c|} 
 \hline
 Link $\{i,j\}$ & Duty cycle (\%) & Link $\{i,j\}$ & Duty cycle (\%) \\ 
 \hline\hline
 $\{13,14\}$ & $49$ \% & $\{6,34\}$ & $44.78$ \% \\ 
 \hline
$\{34,39\}$ & $55.96$ \% & $\{9,26\}$ & $47.3$ \% \\ 
 \hline
$\{9,21\}$ & $52.76$ \% & $\{33,38\}$ & $58.96$ \% \\
 \hline
\end{tabular}
\label{table:1}
\end{table}

\begin{figure}
	\centering
		\includegraphics[width=1\columnwidth]{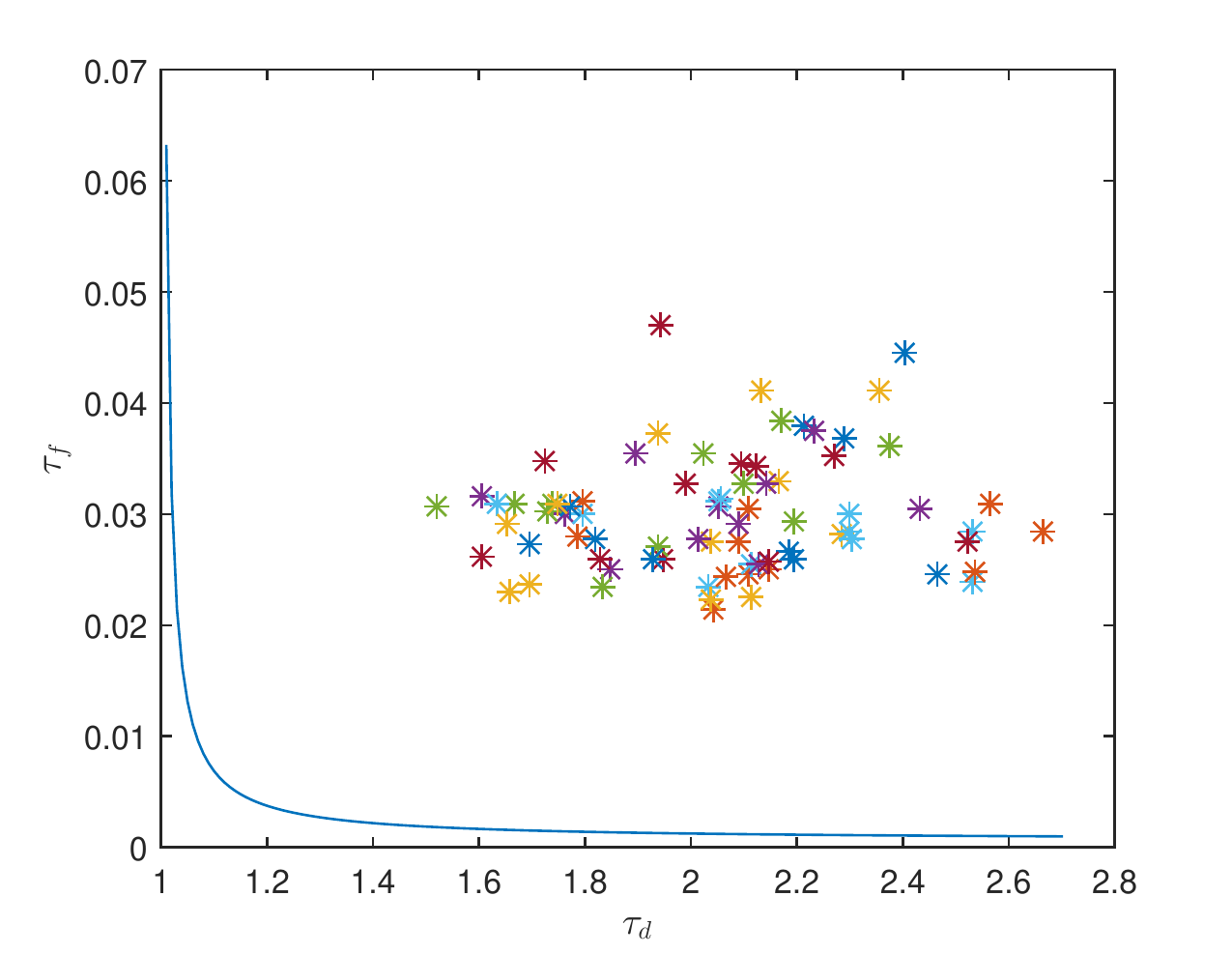}
	\caption{Locus of the points 
	$1/\tau_d+{\Delta_*}/{\tau^{ij}_f}=1$ as a function 
	of $(\tau_d,\tau_f)$ with $\Delta_* = X$ (blue solid line). Notice that
	$\Delta_* = \Delta^{ij}_*$ for all $\{i,j\} \in \E$, so that 
	the locus of point does not vary with $\{i,j\}$. The various $\ast$ represent 
	the values of $(\tau^{ij}_d,\tau^{ij}_f)$ for the network links.}
	\label{fig:fig2}
	\end{figure}
	

\section{Concluding remarks}

We investigated self-triggered coordination for distributed network
systems in the presence of Denial-of-Service at the communication links,
of both genuine and malicious nature. We considered a general framework 
in which DoS can affect each of the network links independently, 
which is relevant for networks operating in 
peer-to-peer mode.
By introducing a notion of Persistency-of-Communication (PoC), we provided an explicit characterization of DoS frequency and duration under which consensus can be preserved by suitably designing time-varying control and communication policies. An explicit characterization of the effects of DoS on the consensus time has also been provided. We compared the 
notion of PoC with classic average connectivity conditions that are found in pure continuous-time consensus networks. The analysis reveals that PoC naturally extends such 
classic conditions to a digital networked setting by requiring graph 
connectivity over periods of time that are consistent with the constraints imposed by the communication medium.  

The present results lend themselves to many extensions. 
Most notably, it is interesting to investigate whether the present
results can be extended to coordination problems involving 
higher-order nodes dynamics.  Another interesting investigation 
pertains the analysis of coordination schemes in the presence 
of both DoS and deceptive attacks.


\appendix


\emph{Proof of Proposition~\ref{prop:persist-synchDoS}.}
Consider any link $\{i,j\} \in \E$, and suppose that
a certain transmission attempt $t^{ij}_k$ is unsuccessful.
We claim that a successful transmission over $\{i,j\}$ does always occur 
within $[t^{ij}_k,t^{ij}_k+\Phi^{ij}]$. 
We prove the claim by contradiction. 
To this end, we first introduce some auxiliary quantities.
Let $\bar H^{ij}_n :=\{h^{ij}_n\} \cup [h^{ij}_n, h^{ij}_n+\tau^{ij} _n+\Delta^{ij}_*[$.
denote the $n$-th DoS interval over the link $\{i,j\}$ prolonged by $\Delta^{ij}_*$ 
units of time. Also let
\begin{eqnarray} \label{eq:Xibar}
\bar \Xi^{ij} (\tau ,t) :=\mathop  \bigcup \limits_{n\in{\integernonnegative}} 
\bar H^{ij}_n \bigcap {[\tau,t ]} \\ 
\bar \Theta^{ij} (\tau ,t) := [\tau,t ] \;  \backslash \; \bar \Xi^{ij} (\tau ,t) 
\end{eqnarray}
Suppose then that the claim is false,
and let $t_*$ denote the last 
transmission attempt over $[t^{ij}_k,t^{ij}_k+\Phi^{ij}]$.
Notice that this necessarily implies $|\bar \Theta^{ij} (t^{ij}_k ,t_{*})|=0$.
To see this, first note that, in accordance with (\ref{eq:st}),
the inter-sampling time over the interval $[t^{ij}_k,t_*]$
is equal to $\varepsilon/(2(d^i+d^j)) = \Delta_*^{ij}$.
Hence, we cannot have $|\bar \Theta^{ij} (t^{ij}_k ,t_{*})|>0$ since this 
would imply the existence of a DoS-free interval within $[t^{ij}_k ,t_{*}]$
of length greater than $\Delta_*^{ij}$, which is not possible since, by hypothesis,
no successful transmission attempt occurs within $[t^{ij}_k ,t_{*}]$.
Thus $|\bar \Theta^{ij} (t^{ij}_k ,t_{*})|=0$.
Moreover, since $t_*$ is unsuccessful, it must be contained 
in a DoS interval, say $H^{ij}_q$. 
This implies $[t_{*},t_{*}+\Delta^{ij}_*[ \subseteq \bar H^{ij}_q$ Hence,
{\setlength\arraycolsep{2pt}
\begin{eqnarray} 
 |\bar \Theta(t^{ij}_k,t_{*}+\Delta^{ij}_*)| &=&  |\bar \Theta(t^{ij}_k,t_{*})| + |\bar \Theta(t_{*},t_{*}+\Delta^{ij}_*)| 
 \nonumber \\ &=& 0
\end{eqnarray}}%
However, condition $|\bar \Theta(t^{ij}_k,t_{*}+\Delta^{ij}_*)| =0$ 
is not possible. To see this, simply notice that
{\setlength\arraycolsep{2pt}
\begin{eqnarray} \label{eq:Lemma1}
|\bar \Theta(t^{ij}_k,t)| &=& t-t^{ij}_k - |\bar \Xi(t^{ij}_k,t)| \nonumber \\
&\geq& t-t^{ij}_k - |\Xi(t^{ij}_k,t)| - (n(t^{ij}_k,t) + 1)  \Delta^{ij}_*  \nonumber \\
&\geq&  ( t-t^{ij}_k ) ( 1 - \alpha^{ij} ) -\kappa^{ij} - (\eta^{ij} +1) \Delta^{ij}_*  
\end{eqnarray}}%
for all $t \geq t^{ij}_k$
where the first inequality follows from the definition 
of the set $\bar \Xi(\tau,t)$ while the second one 
follows from Assumption 1 and 2. 
Hence, by (\ref{eq:Lemma1}), we have
$|\bar \Theta(t^{ij}_k,t)|>0$ for all
$t>t^{ij}_k + ( 1 - \alpha^{ij} )^{-1} (\kappa^{ij} + (\eta^{ij} +1)\Delta^{ij}_*) = t^{ij}_k + \Phi^{ij}$.
Accordingly, $|\bar \Theta(t^{ij}_k,t_{*}+\Delta^{ij}_*)| =0$
cannot occur because $t_{*}+\Delta^{ij}_* >  t^{ij}_k + \Phi^{ij}$.
In fact, by hypothesis, $t_*$ is defined as the last 
unsuccessful transmission attempt within $[t^{ij}_k,t^{ij}_k+\Phi^{ij}]$,
and, by (\ref{eq:st}), the next transmission attempt after $t_*$ occurs 
at time $t_{*}+\Delta^{ij}_*$. This concludes the proof. \qedp

\ifCLASSOPTIONcaptionsoff
  \newpage
\fi

\end{document}